\documentclass[runningheads]{llncs}

\usepackage{url}
\usepackage{graphicx}
\usepackage{amssymb,amsmath,enumerate}
\usepackage{afterpage}
\usepackage{textcomp}
\RequirePackage{amsmath} 
\RequirePackage{xspace}
\usepackage{xspace}
\usepackage{bbm}
\usepackage{xcolor}
\usepackage{soul}
\usepackage{booktabs}
\usepackage{multirow}
\usepackage{hyperref}
\usepackage{pdfpages}

\usepackage{tabularx}

\def\XS{\xspace}
\DeclareMathAlphabet{\mathb}{OML}{cmm}{b}{it}
\def\sbm#1{\ensuremath{\mathb{#1}}}                % Style gras italique (necessite amsmath)
\def\sbmm#1{\ensuremath{\boldsymbol{#1}}}          % Style gras italique (necessite amsmath)
\def\Ab{{\sbm{A}}\XS}

\def\Db{{\sbm{D}}\XS}  \def\db{{\sbm{d}}\XS}

\def\Sb{{\sbm{S}}\XS}  \def\sb{{\sbm{s}}\XS}

  \def\wb{{\sbm{w}}\XS}

      \def\Deltab    {{\sbmm{\Delta}}\XS}

\def\thetab      {{\sbmm{\theta}}\XS}      \def\Thetab    {{\sbmm{\Theta}}\XS}

\def\mub         {{\sbmm{\mu}}\XS}
\def\nub         {{\sbmm{\nu}}\XS}

\def\phib        {{\sbmm{\phi}}\XS}

\def\PM{\kern0pt^{\textrm{{\scriptsize PM}}}\kern0pt}
\def\MMAP{\kern1pt^{\textrm{{\tiny MMAP}}}\kern-1pt}

\newcommand{\Wv}{\ensuremath{\mathbf{W}}} % Vecteur des var observees.(obs)
\newcommand{\wv}{\ensuremath{\mathbf{w}}} % Vecteur des var observees.(obs)
\newcommand{\Xv}{\ensuremath{\mathbf{X}}} % Vecteur des var observees.(obs)
\newcommand{\xv}{\ensuremath{\mathbf{x}}} % Vecteur des var observees.(obs)
\newcommand{\Yv}{\ensuremath{\mathbf{Y}}} % Vecteur des var observees.(obs)
\newcommand{\yv}{\ensuremath{\mathbf{y}}} % Vecteur des var observees.(obs)
\def\wp1{\mathrm{w.p.} 1}  % with respect to
\begin{document}

\title{Towards frugal unsupervised detection of subtle abnormalities in medical imaging}

\titlerunning{Towards frugal UAD in medical imaging}

\author{Geoffroy Oudoumanessah\inst{1,2,3} \and
Carole Lartizien \inst{3} \and
Michel Dojat \inst{2} \and Florence Forbes \inst{1}} 
%index{Oudoumanessah, Geoffroy}
%index{Lartizien, Carole}
%index{Dojat, Michel}
%index{Forbes, Florence}

\authorrunning{G. Oudoumanessah et al.}
\institute{
Univ. Grenoble Alpes, Inria, CNRS, Grenoble INP, LJK, 38000 Grenoble, France \\
\email{\{first.last\}@inria.fr} \and
Univ. Grenoble Alpes, Inserm U1216, CHU Grenoble Alpes, Grenoble Institut des Neurosciences, 38000 Grenoble, France \\
\email{\{first.last\}@univ-grenoble-alpes.fr} \and
Univ. Lyon, CNRS, Inserm, INSA Lyon, UCBL, CREATIS, UMR5220, U1294, F‐69621, Villeurbanne, France\\ 
\email{\{first.last\}@creatis.insa-lyon.fr}}
\maketitle              

\begin{abstract}
Anomaly detection in medical imaging is a challenging task in  contexts where abnormalities are not annotated. This problem can be addressed through unsupervised anomaly detection (UAD) methods, which identify features that do not match with a reference model of normal profiles.
Artificial neural networks  have been extensively used for UAD but they do not generally achieve an optimal trade-off between accuracy and computational demand. As an alternative, we investigate mixtures of probability distributions whose versatility has been widely recognized for a variety of data and tasks, while not requiring excessive design effort or  tuning. Their expressivity makes them good candidates to account for complex multivariate reference models. Their much smaller number of parameters makes them more amenable to interpretation and efficient learning. However, standard estimation procedures, such as the Expectation-Maximization algorithm, do not scale well to large data volumes as they require high memory usage. To address this issue, we propose to incrementally compute inferential quantities. 
This online approach is illustrated on the challenging detection of subtle abnormalities in MR brain scans for the follow-up of newly diagnosed Parkinsonian patients. The identified structural abnormalities are consistent with the disease progression, as accounted by the Hoehn and Yahr scale.


\end{abstract}

\section{Introduction}

Despite raising concerns about the environmental impact of artificial intelligence \cite{Schwartz2020,Strubell2019,thompson2022computational}, the question of resource efficiency has not yet really reached medical imaging studies. The issue has multiple dimensions and  the lack of clear metrics for a fair  assessment of algorithms,  in terms of resource and energy consumption, contrasts with the obvious healthcare benefits of the ever growing performance of machine and statistical learning solutions. 

In this work, we investigate the case of subtle abnormality detection in medical images, in an unsupervised context usually referred to as \textit{Unsupervised Anomaly Detection} (UAD). 
This formalism requires only the identification of {\it normal} data to construct a normative model. \textit{Anomalies} are then detected as outliers, {\it i.e.} as samples deviating from this normative model. 
Artificial neural networks (ANN) have been extensively used for UAD \cite{UPD_study}. Either based on standard autoencoder (AE) architectures \cite{baur_autoencoders_2020} or on more advanced architectures, {\it e.g.} combining a vector quantized AE with autoregressive transformers \cite{pinaya_MEDIA22}, ANN do not generally achieve an optimal trade-off between accuracy and computational demand. 
As an alternative, we show that  more {\it frugal} approaches can be reached with traditional statistical models 
provided their cost in terms of memory usage can be addressed. 
Frugal solutions usually refer to  strategies that can run with limited 
resources such as that of a single laptop. 
Frugal learning has been studied from several angles,
in the form of constraints on the data acquired, on the algorithm deployed and on the nature of the proposed solution \cite{Evchenko2021}. The angle we adopt is that of
{\it online} or incremental learning, which refers to approaches that handle data  in a sequential manner resulting in more efficient solutions in terms of memory usage and overall energy consumption.
For UAD, we propose to investigate
mixtures of probability distributions whose interpretability and versatility have been widely recognized for a variety of data and tasks, while not requiring excessive design effort or  tuning. In particular, the use of multivariate Gaussian or generalized Student mixtures has been already demonstrated in many anomaly detection tasks, see \cite{Arnaud2018,Munoz2019,OLUWASEGUN2023} and references therein or \cite{UPD_study} for a more general recent review. 
However, in their standard {\it batch} setting, mixtures are difficult to use with 
huge datasets due to the  dramatic increase of time and memory consumption
required by their estimation traditionally performed with an  Expectation-Maximization (EM) algorithm \cite{McLachlan2007}. 
Online more tractable versions of EM have been proposed and theoretically studied in the literature, {\it e.g.} \cite{Cappe2009,Fort2021}, but with some restrictions on the class of mixtures that can be handled this way.
A first natural approach is to consider Gaussian mixtures that belong to this class.
We thus, present improvements regarding the implementation of  an online EM for  Gaussian mixtures. We then consider  more general mixtures based on {\it multiple scale t-distributions} (MST) specifically adapted to outlier detection \cite{ForbesWraith14}.  We show that these mixtures  can be cast into the online EM framework and describe the resulting algorithm. 

Our approach is illustrated with  the 
MR imaging exploration of \textit{de novo} (just diagnosed) Parkinson's Disease (PD) patients, where brain anomalies are subtle and hardly visible in standard T1-weighted or diffusion MR images. 
The anomalies detected by our method are consistent with the Hoehn and Yahr (HY) scale \cite{HY}, which describes how the symptoms of Parkinson's disease progress. The results provide additional interesting clinical insights by pointing out the most impacted subcortical structures at both HY stages 1 and  2. The use of such an external scale  appears to be an original and relevant indirect validation, in the absence of ground truth at the voxel level. 
Energy and memory consumptions are also reported for batch and online EM to confirm the interesting performance/cost trade-off achieved. The code is available at \href{https://github.com/geoffroyO/onlineEM}{https://github.com/geoffroyO/onlineEM}.

\section{UAD with mixture models}
\label{sec:uad}

Recent studies  have shown that, on subtle lesion detection tasks with limited data, alternative approaches to ANN, such as {\it one class support vector machine} or mixture models \cite{Arnaud2018,Munoz2019}, were performing similarly \cite{OLUWASEGUN2023,pinon2023}. 
We further investigate mixture-based models and show how the main UAD steps, {\it i.e.} the construction of a reference model and of a decision rule, can be designed. %\cite{ElAzami_PlosOne2016}

\noindent {\bfseries Learning a reference model.} 
We  consider a set $\mathbb{Y}_H$ of voxel-based features for a number of control ({\it e.g.} healthy) subjects, $\mathbb{Y}_H=\{\yv_v,v \in \mathbb{V}_H\}$ 
where $\mathbb{V}_H$  represents the voxels of all control subjects and $\yv_v \in \bbbr^M$ is typically deduced from image modality maps at voxel $v$ or from abstract representation features provided by some ANN performing a pre-text task \cite{Li2021}. 
To account for the distribution of such normal feature vectors, we consider two types of  mixture models, mixtures of Gaussian distributions with high tractability in multiple dimensions and mixtures of multiple scale t-distributions (MST) that are more appropriate when the data present elongated and strongly non-elliptical subgroups
\cite{ForbesWraith14,Arnaud2018,Munoz2019}. 
By fitting such a mixture model to the control data $\mathbb{Y}_H$, we build a reference model density $f_H$ that depends on some parameter $\Thetab_H = \{\thetab_k, \pi_k, k=1:K_H\}$:
\vspace*{-0.2cm}
\small
\begin{eqnarray}
f_H(\yv;\Thetab_H)&=&\sum_{k=1}^{K_H}\pi_k f(\yv;\thetab_k), \label{def:fH}
\end{eqnarray}
\normalsize
with $\pi_k \!\in \![0,1]$, $\sum_{k=1:K_H} \pi_k \!= \!1$
and $K_H$ the number of  components,  each  characterized by a distribution $f(\cdot; \thetab_k)$. The EM algorithm is usually used to estimate $\Thetab_H$ that  best fits $\mathbb{Y}_H$ while  $K_H$ can be estimated using {\it the slope heuristic} \cite{BaudryMaugisMichel2012}. 

\noindent{\bfseries Designing a  proximity measure.}
Given  a reference model (\ref{def:fH}), a measure of proximity $r(\yv_v; \Thetab_H)$ of voxel $v$ (with value $\yv_v$)  to $f_H$ needs to be chosen.
To make use of the mixture structure, we propose to consider  distances to the respective mixture components through some weights acting as inverse Mahalanobis distances. We specify below this new proximity measure for  MST mixtures. 
 MST distributions  are  generalizations of the multivariate t-distribution that extend its Gaussian scale mixture representation \cite{Kotz2004}. The standard t-distribution univariate scale (weight) variable is replaced by a $M$-dimensional scale (weight) variable  $\Wv=(W_m)_{m=1:M} \in \mathbb{R}^M$ with $M$  the features dimension,
\begin{gather}
\label{eq:MST}
\scalebox{1}{
$f_{\mathcal{MST}}(\mathbf{y} ; \boldsymbol{\theta}) = \int \limits_{[0,\infty]^M} \mathcal{N}_M(\mathbf{y}; \boldsymbol{\mu}, \boldsymbol{D\Delta_wAD}^T) \prod \limits_{m=1}^M \mathcal{G}\left ( w_m; \frac{\nu_m}{2} \right)dw_{1}\ldots dw_M$}, 
\end{gather}
where $\mathcal{G}(\cdot , \frac{\nu_m}{2})$  denotes the gamma  density with parameter  $(\frac{\nu_m}{2}, \frac{\nu_m}{2}) \in \mathbb{R}^2$ and $\mathcal{N}_M$ the multivariate normal distribution with mean parameter $\boldsymbol{\mu} \in \mathbb{R}^M$ and 
covariance matrix $\mathbf{D \Delta_w AD}^T$ showing the scaling by the $W_m$'s through a diagonal matrix $\boldsymbol{\Delta}_w = diag(w_1^{-1}, \ldots, w_M^{-1})$. 
The MST parametrization uses the spectral decomposition of the scaling matrix $\boldsymbol{\Sigma}=\mathbf{DAD}^T$, with $\mathbf{D} \in \mathcal{O}(M) \subset \mathbb{R}^{M\times M}$ orthogonal and $\mathbf{A}=diag(A_1, \ldots, A_M)$ diagonal. The whole set of parameters is
$\boldsymbol{\theta} = \{\boldsymbol{\mu}, \boldsymbol{A}, \boldsymbol{D}, (\nu_m)_{m=1:M}\}$.
The scale variable $W_m$ for dimension $m$ can be interpreted as accounting for the weight of this dimension and can be used to derive a measure of proximity. After fitting a mixture (\ref{def:fH}) with MST components to $\mathbb{Y}_H$, we set $r(\mathbf{y}_v;\boldsymbol{\Theta}_H ) = \max_{m=1:M} \Bar{w}_m^{\mathbf{y}_v}$, with $\Bar{w}_m^{\mathbf{y}} = \mathbb{E}[W_m | \mathbf{y}; {\boldsymbol{\Theta}}_H]$. The proximity $r$  is typically larger when at least one dimension of $\mathbf{y}_v$ is well explained by the model. A similar proximity measure can also be derived for Gaussian mixtures, see details in the Supplementary Material  Section 1.

\noindent{\bfseries Decision rule.}
For an effective   detection, a threshold $\tau_\alpha$ on proximity scores can be computed in a  data-driven way by deciding on an acceptable false positive rate (FPR) $\alpha$; $\tau_\alpha$ is the value such that $P(r(\Yv; \Thetab_H) < \tau_\alpha) = \alpha, $ when $\Yv$ follows the $f_H$ reference distribution. All  voxels $v$ whose proximity $r(\yv_v ; \Thetab_H)$ is below $\tau_\alpha$ are then labeled as abnormal. In practice, while $f_H$ is known explicitly, the probability distribution of $r(\Yv; \Thetab_H)$ is not. However, it is easy to simulate this distribution or to estimate  $\tau_\alpha$ as an empirical $\alpha$-quantile \cite{Arnaud2018}. Unfortunately, learning  $f_H$ on huge datasets may not be possible  due to the dramatic increase in time, memory and energy required by the EM algorithm. This issue often arises in medical imaging  with the increased availability of multiple 3D modalities as well as the emergence of image-derived parametric maps such as radiomics \cite{Gillies} that should be analysed jointly, at the voxel level, and for a large number of subjects. 
A possible solution consists of employing powerful computers with graphics cards or grid-architectures in cloud computing. Here, we show that a more resource-friendly solution is possible using an online version of  EM detailed in the next section.

\section{Online mixture learning for large data volumes}
\label{sec:online}

Online learning refers to procedures able to deal with data acquired sequentially. Online  variants of EM, among others, are described in \cite{Cappe2009,Maire:2017aa,Karimi:2019ab,Karimi:2019aa,Fort:2020aa,Kuhn:2020aa,Nguyen:2020ac}. As an archetype of such algorithms, we  consider the online EM of \cite{Cappe2009} which belongs to 
 the family of stochastic approximation  algorithms \cite{Borkar2008}. 
This algorithm has been well theoretically studied and extended. However, it is designed only for distributions that admit a data augmentation scheme yielding a  complete likelihood of the exponential family form, see (\ref{eq: A1}) below.  This case is already very broad, including Gaussian, gamma, t-distributions, etc. and mixtures of those.  We recall below the main assumptions required and the online EM iteration. 

Assume $\left(\Yv_{i}\right)_{i=1}^{n}$ is a sequence of 
 $n$ independent and identically distributed  replicates of a random  variable $\Yv\!\in\! \mathbb{Y}\!\subset\!\bbbr^{M}$, observed 
one at a time. Extension to successive mini-batches of observations is  straightforward \cite{Nguyen:2020ac}.
In addition, $\Yv$ is assumed to be the visible part of the pair $\Xv^{\top}\!=\!\left(\Yv^{\top},\sbm{Z}^{\top}\right) \!\in\mathbb{X} $,
where $\mathbf{Z}\in\bbbr^{l}$ is a latent variable, {\it e.g.} the unknown component label in a mixture model, and $l\in\bbbn$. That is, each $\Yv_{i}$ 
is the visible part of a pair $\Xv_{i}^{\top}\!=\!\left(\Yv_{i}^{\top},\mathbf{Z}_{i}^{\top}\right)$.
Suppose $\Yv$ arises from some data generating process (DGP)
 characterised by a probability density function  $f\left(\yv;\sbmm{\theta}_0\right)$,
with unknown parameters $\boldsymbol{\theta}_0\in\mathbb{T}\subseteq\bbbr^{p}$,
for $p\in\bbbn$.

Using the sequence $\left(\Yv_{i}\right)_{i=1}^{n}$, the  method of \cite{Cappe2009}
sequentially estimates  $\boldsymbol{\theta}_{0}$ provided the  following assumptions are met:

\begin{itemize}
\item [{(A1)}] The complete-data likelihood for $\Xv$
is of the exponential family form: 
\begin{equation}
f_{c}\left(\xv;\boldsymbol{\theta}\right)=h\left(\xv\right)\exp\left\{ \left[\mathbf{s}\left(\xv\right)\right]^{\top}\boldsymbol{\phi}\left(\boldsymbol{\theta}\right)-\boldsymbol{\psi}\left(\boldsymbol{\theta}\right)\right\} \text{,}\label{eq: A1}
\end{equation}
with $h\!:\!\bbbr^{M+l}\!\rightarrow\!\left[0,\infty\right)$, $\psi\!:\!\bbbr^{p}\!\rightarrow\!\bbbr$,
$\mathbf{s}\!:\!\bbbr^{M+l}\!\rightarrow\!\bbbr^{q}$, $\boldsymbol{\phi}\!:\!\bbbr^{p}\rightarrow\bbbr^{q}$,
for $q\in\bbbn$.
\item [{(A2)}] The function
\begin{align}
\bar{\mathbf{s}}\left(\yv;\boldsymbol{\theta}\right) & =\mathbb{E}\left[\mathbf{s}\left(\Xv\right)|\Yv=\yv ; \boldsymbol{\theta}\right]\label{eq: A2}
\end{align}
is well-defined for all $\yv$ and $\boldsymbol{\theta}\in\mathbb{T}$,
where $\mathbb{E}\left[\cdot|\Yv=\yv;\boldsymbol{\theta}\right]$
is the conditional expectation when $\Xv$
arises from the DGP characterised by $\boldsymbol{\theta}$.
\item [{(A3)}] There is a convex $\mathbb{S}\subseteq\bbbr^{q}$,
 satisfying:
(i) for all $\gamma\in\left(0,1\right)$, $\mathbf{s}\in\mathbb{S}$, $\yv\in\mathbb{Y}$, and
$\boldsymbol{\theta}\in\mathbb{T}$,
$\left(1-\gamma\right)\mathbf{s}+\gamma\bar{\mathbf{s}}\left(\yv;\boldsymbol{\theta}\right)\in\mathbb{S}\text{;}$ and 
(ii) for any $\mathbf{s}\in\mathbb{S}$, the function 
$Q\left(\mathbf{s};\boldsymbol{\theta}\right)=\mathbf{s}^{\top}\boldsymbol{\phi}\left(\boldsymbol{\theta}\right)-\psi\left(\boldsymbol{\theta}\right)$
has a unique global maximizer on $\mathbb{T}$ denoted by
\begin{equation}
\bar{\boldsymbol{\theta}}\left(\mathbf{s}\right)=\underset{\boldsymbol{\theta}\in\mathbb{T}}{\arg\max}\;Q\left(\mathbf{s};\boldsymbol{\theta}\right)\text{.}\label{eq: def theta}
\end{equation}
\end{itemize}
Let $\left(\gamma_{i}\right)_{i=1}^{n}$ be a sequence of learning
rates in $\left(0,1\right)$ and let $\boldsymbol{\theta}^{\left(0\right)}\in\mathbb{T}$
be an initial estimate of $\boldsymbol{\theta}_{0}$. For each $i=1:n$,
the online EM of \cite{Cappe2009} proceeds by computing 
\begin{equation}
\mathbf{s}^{\left(i\right)}=\gamma_{i}\bar{\mathbf{s}}(\yv_{i};\boldsymbol{\theta}^{(i-1)})+\left(1-\gamma_{i}\right)\mathbf{s}^{\left(i-1\right)}\text{,}\label{eq: S up}
\end{equation}
and
\begin{equation}
\boldsymbol{\theta}^{\left(i\right)}=\bar{\boldsymbol{\theta}}(\mathbf{s}^{\left(i\right)})\text{,}\label{eq: Theta up}
\end{equation}
where $\mathbf{s}^{\left(0\right)}=\bar{\mathbf{s}}(\yv_{1};\boldsymbol{\theta}^{\left(0\right)})$.
It is shown in Thm. 1 of \cite{Cappe2009} that when $n$ tends to infinity, the sequence 
$(\boldsymbol{\theta}^{\left(i\right)})_{i=1:n}$ 
of estimators of $\boldsymbol{\theta}_{0}$  satisfies a convergence result to stationary points of the likelihood (cf.  \cite{Cappe2009} for a more precise statement). 

In practice, the algorithm implementation requires two  quantities,  $\bar{\sb}$ in (\ref{eq: A2}) and $\bar{\thetab}$ in (\ref{eq: def theta}). They are necessary to define the updating of sequences $(\mathbf{s}^{\left(i\right)})_{i=1:\infty}$ and
$(\boldsymbol{\theta}^{\left(i\right)})_{i=1:\infty}$. 
We detail below these quantities for a MST mixture.

\noindent{\bfseries Online MST mixture EM.}
As shown in \cite{NguyenForbes2021}, the mixture case can be deduced from a single component case. The  exponential form
for a MST (\ref{eq:MST}) writes:
\begin{eqnarray}
f_c(\xv; \thetab) &= & {\cal N}_M(\yv; \mub,\Db \Deltab_\wv\Ab \Db^T) \prod_{m=1}^M {\cal G}\left(w_m; \frac{\nu_m}{2}\right), \quad \mbox{with } \xv = (\yv, \wv)  \label{def:fcmst} \; \\
&=&h(\yv,\wb) \exp\left([\sb(\yv,\wb)]^T \phib(\mub,\Db,\Ab,\nub) - \psi(\mub,\Db,\Ab,\nub)\right) \nonumber
\end{eqnarray}
with
$
\sb(\yv,\wb)\! =\! \left[
 w_1\yv, 
  w_1 vec(\yv\yv^\top\!),
  w_1,
   \log \!w_1,
   \ldots,
   w_M\yv,
    w_M vec(\yv\yv^\top\!),
     w_M,
     \log\! w_M \!\right]^\top\!\!, $ 
     $\phib(\mub,\Db,\Ab,\nub)= [\phib_1, \ldots, \phib_M]^T$
     with $\phib_m$ equal to:
\begin{eqnarray*}
&&\phib_m = \left[\displaystyle
 \frac{\db_m\db_m^T \mub}{A_m},
 \displaystyle  -\frac{vec(\db_m\db_m^T)}{2 A_m},
  \displaystyle  -\frac{vec(\db_m\db_m^T)^T vec(\mub\mub^T)}{2 A_m} -\frac{\nu_m}{2},
  \displaystyle \frac{1+\nu_m}{2}\right]\\
\mbox{and} && \psi(\mub,\Db,\Ab,\nub)  = \sum_{m=1}^M \left(\frac{\log A_m}{2} + \log \Gamma(\frac{\nu_m}{2}) - \frac{\nu_m}{2} 
\log(\frac{\nu_m}{2}) \right) \; , 
\end{eqnarray*}
where $\db_m$ denotes the $m^{th}$ column of $\Db$ and  $vec(\cdot)$ the vectorisation operator, which converts a  matrix to a column vector. 
The exact form of $h$ is not important for the algorithm. 
It follows that $\bar{\thetab}(\sb)$ is defined as the unique maximizer of function $Q(\sb, \thetab) = \sb^T\phib(\thetab) - \psi(\thetab)$ where $\sb$ is a vector that matches the definition and dimension 
of $\phib(\thetab)$ and can be conveniently written as
$\sb = [
 \sb_{11},
 vec(\Sb_{21}),
 s_{31},
 s_{41},
  \ldots,
     \sb_{1M},
      vec(\Sb_{2M}),
      s_{3M},
      s_{4M}]^T,$
with for each $m$, 
$\sb_{1m}$ is a $M$-dimensional vector, $\Sb_{2m}$ is a $M\times M$ matrix, $s_{3m}$ and $s_{4m}$ are scalars. 
Solving for the roots of the $Q$ gradients  leads to $\bar{\thetab}(\sb)= (\bar{\mub}(\sb), \bar{\Ab}(\sb), \bar{\Db}(\sb), \bar{\nub}(\sb))$ whose expressions are detailed in Supplementary Material Section 2. 

 A second important quantity is $\bar{\sb}(\yv, \thetab) = \mathbb{E}\left[\mathbf{s}\left(\Xv\right)|\Yv=\yv; \boldsymbol{\theta}\right]$. This quantity requires to compute the following expectations for all $m$, $ \mathbb{E}\left[W_m |\Yv=\yv;\boldsymbol{\theta} \right]$ and  $ \mathbb{E}\left[\log W_m |\Yv=\yv; \boldsymbol{\theta}\right]$. 
 More specifically in the update equation (\ref{eq: S up}), these expectations need to be computed for $\yv=\yv_i$ the observation at iteration $i$. We therefore denote these expectations respectively by
 \begin{equation}
      u_{im}^{(i-1)} =  \mathbb{E}[W_m |\Yv=\yv_i; \boldsymbol{\theta}^{(i-1)}] =\alpha_m^{(i-1)}/\beta_m^{(i-1)}
 \end{equation}

 and 
 $  \tilde{u}_{im}^{(i-1)} =  \mathbb{E}[\log W_m |\Yv=\yv_i; \boldsymbol{\theta}^{(i-1)}] =  \Psi^{(0)}(\alpha_m^{(i-1)}) - \log \beta_m^{(i-1)} \; ,$
 where 
 $\alpha_m^{(i-1)} = \frac{\nu_m^{(i-1)} + 1}{2}$ and 
 $\beta_m^{(i-1)}  =  \frac{\nu_m^{(i-1)}}{2} + \frac{ \left(\db_m^{(i-1)T}(\yv_i - \mub^{(i-1)})\right)^2}{2A^{(i-1)}_m}\; . $
 The update of $ \mathbf{s}^{(i)}$ in (\ref{eq: S up}) follows from the update for each $m$. From this single MST iteration, the mixture case is easily derived, see \cite{NguyenForbes2021} or Supplementary Material  Section 2. 

 \noindent{\bfseries Online Gaussian mixture EM.} This case can be found in previous work {\it e.g.} \cite{Cappe2009,Nguyen:2020ac} but to our knowledge, implementation optimizations are never really addressed. 
 We propose an original version that  saves computations, especially in a multivariate case where  $\bar{\boldsymbol{\theta}}\left(\mathbf{s}\right)$ 
   involves large matrix inverses and determinants. Such inversions are avoided using results detailed in Supplementary  Section 3.

\section{Brain abnormality exploration in \textit{de novo} PD patients}

\noindent{\bfseries Data description and preprocessing.}
The Parkinson's Progression Markers Initiative (PPMI) \cite{ppmi}  is an open-access database dedicated to PD. It includes MR images of \textit{de novo} PD patients, as well as of healthy subjects (HC), all acquired on the same 3T Siemens Trio Tim scanner. For our illustration, we use 108 HC and 419 PD samples, each composed of a 3D T1-weighted image (T1w), Fractional Anisotropy (FA) and Mean Diffusivity (MD) volumes. The two latter are extracted from diffusion imaging using the DiPy  package \cite{dipy}, registered onto T1w and interpolated to the same spatial resolution with SPM12.  Standard T1w preprocessing steps, comprising non-local mean denoising, skull stripping and tissue segmentation are also performed with  SPM12. HC and PD groups are age-matched (median age: 64 y.) with the male-female ratio equal to 6:4. We focus on some subcortical structures, which are mostly impacted at the early stage of the disease \cite{Dexter},  Globus Pallidus external and internal (GPe and GPi), Nucleus Accumbens (NAC), Substantia Nigra reticulata (SNr), Putamen (Pu), Caudate (Ca) and Extended Amygdala (EXA). Their position is determined by projecting the CIT168 atlas \cite{pauli2018} onto each individual image.

\noindent{\bfseries Pipeline and results.} We follow Sections \ref{sec:uad} and \ref{sec:online}
using  T1w, FA and MD volumes as features ($M =3$) and a FPR $\alpha=0.02$. 
The pipeline is repeated 10 times for cross-validation. Each fold is composed of 64 randomly selected HC images for training  (about 70M voxels), the remaining 44 HC and all the PD samples for testing.
For the reference model, we test Gaussian and MST mixtures, with respectively $K_H=14$ and $K_H=8$, estimated with the slope heuristic.
Abnormal voxels are then detected for all test subjects, on the basis of their proximity to the learned reference model, as detailed in Section 2.

The PPMI does not provide ground truth information at the voxel level. This is a recurring issue in UAD,  which limits validations to mainly qualitative ones. For a more quantitative evaluation, we propose to resort to an auxiliary task whose success is likely to be correlated with a good anomaly detection. We consider the classification of test subjects into healthy and Parkinsonian subjects based on their global (over all brain) percentages of abnormal voxels. We exploit the availability of HY values to divide the patients into two HY=1 and HY=2 groups, representing the two early stages of the disease's progression. Classification results yield  a median g-mean, for stage 1 vs stage 2, respectively of 0.59 vs 0.63 for the Gaussian mixtures model and  0.63 vs 0.65 for the MST mixture. 
The ability of both mixtures to better differentiate stage 2 than stage 1 patients from HC is consistent with the progression of the disease. Note that the structural differences between these two PD stages remain subtle and difficult to detect, demonstrating the efficiency of the models. The MST mixture model appears better in identifying stage 2 PD patients based on their abnormal voxels.

\begin{figure}[ht]
  \centering
  \includegraphics[width=\textwidth]{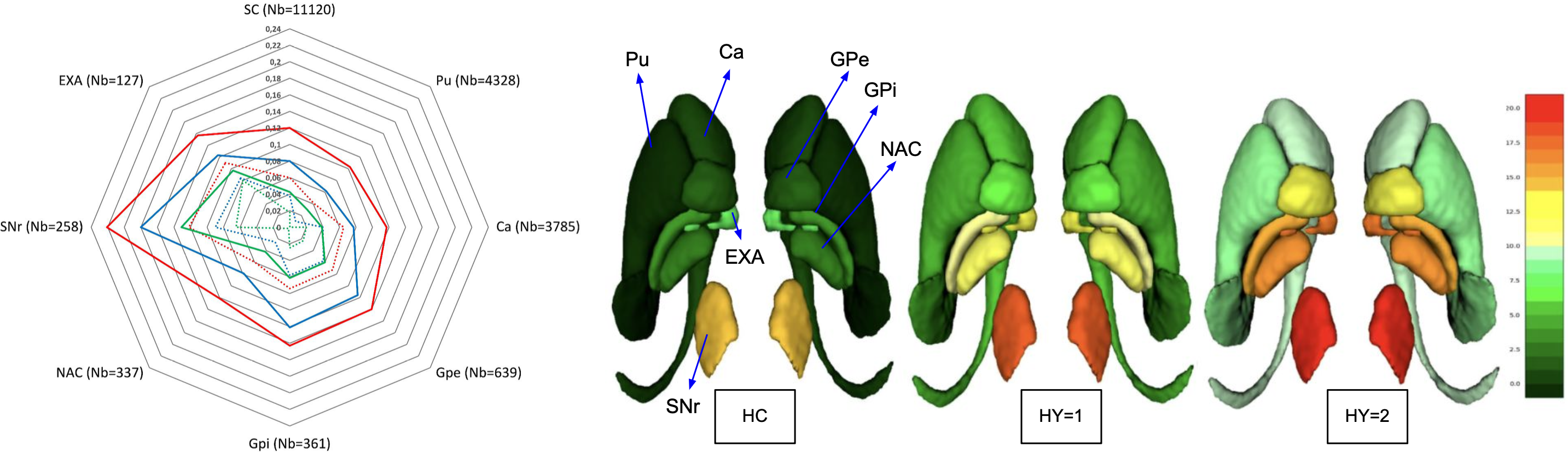}
  \caption{Left: Median, over 10 folds, percentages of anomalies (0 to 22\%) in each subcortical structure (see text for full names) for control subjects (green), stage 1 (blue) and stage 2 (red) patients. Plain and dotted lines indicate respectively results obtained with the MST and Gaussian mixtures. Structure sizes in voxels are indicated in parenthesis. SC refers to the combination of all structures. Right: 3D rendering of the subcortical structures colored according to MST percentages from 0\% (green) to 22\% (red), for healthy controls (HC), stage 1 and stage 2 groups.}
    \label{fig:pc}
\end{figure}

To gain further insights, we report, in Fig.~\ref{fig:pc}, the percentages of anomalies detected in each subcortical structure, for control, stage 1 and stage 2 groups. For each structure and both mixture models, the number of anomalies increases from control to stage 1 and stage 2 groups. As expected the MST mixture shows a better ability to detect outliers with significant differences between HC and PD groups, while for the Gaussian model, percentages do not depart much from that in the control group. Overall, in line with the know pathophysiology \cite{Dexter}, MST results suggest clearly that all structures are potential good markers of the disease progression at these early stages, with GPe, GPi, EXA and SNr showing the largest impact. 

Regarding efficiency, energy consumption in kilojoules (kJ) is measured using the PowerAPI library \cite{powerapi}. In Table~\ref{tab1}, we report the energy consumption for the training and testing of one random fold, comparing our online mixtures with AE-supported methods for UAD \cite{baur_autoencoders_2020}, namely the patch-based reconstruction error \cite{baur_autoencoders_2020} and FastFlow \cite{yu2021fastflow}. We implemented both methods with two different AE architectures: a lightweight AE already used for \textit{de novo} PD detection \cite{pinon2023}, and a larger one, ResNet-18 \cite{he2016deep}. The global g-mean (not taking HY stages into account) is also reported for the chosen fold. The experiments were run on a CPU with Intel Cascade Lake 6248@2.5GHz (20 cores), and a GPU Nvidia V100-32GB. Online mixtures exhibit significantly lower energy consumption, both for training and inference. In terms of memory cost, DRAM peak results, as measured by the \textit{tracemalloc} Python library, also show lower costs for online mixtures, which by design deal with batches of voxels of smaller sizes than the batches of patches used in AE solutions. These results highlight the advantage of online mixtures, which compared to other hardware-demanding methods, can be run on a minimal configuration while maintaining good performance.

\begin{table}[!ht]
  \centering
  \caption{UAD methods comparison for one fold: 
online Gaussian (OGMM) and Student (OMMST) mixtures, Lightweight AE and ResNet-18 architectures with reconstruction error (RE) and FastFlow (FF) based detection. Best values in bold font.}
  \scalebox{.7}{
    \begin{tabular}{c|c|ccc|ccc|cc}
    \multicolumn{2}{r}{} & \multicolumn{3}{c}{Training} & \multicolumn{3}{c}{Inference} & \multicolumn{2}{r}{} \\
    Method & Backend & Time  & Consumption & DRAM peak & Time  & Consumption & DRAM peak & Gmean & Parameters \\
    \midrule
    \multicolumn{10}{c}{Online Mixtures (ours)} \\
    \midrule
    OGMM  & CPU   & \textbf{50s} & \textbf{85 kJ} & \textbf{494 MB} & \textbf{17min} & \textbf{23 kJ} & \textbf{92 MB} & 0.65  & 140 \\
    OMMST & CPU   & 1min20 & 153 kJ & 958 MB & 18min & 32 kJ & 96 MB & \textbf{0.67} & {\bf 128} \\
    \midrule
    \multicolumn{10}{c}{Lightweight AE} \\
    \midrule
    RE    & GPU   & 1h26  & 5040 kJ & 26 GB & 3h30  & 8350 kJ & 22 GB & 0.61  & 5266 \\
    FF    & GPU   & 4h    & 6854 kJ & 27 GB & 3h53  & 13158 kJ & 27 GB & 0.55  & 1520 \\
    \midrule
    \multicolumn{10}{c}{Resnet-18} \\
    \midrule
    RE    & GPU   & 17h40 & 53213 kJ & 26 GB & 59h   & 108593 kJ & 28 GB & 0.64  & 23730218 \\
    FF    & GPU   & 4h10  & 7234 kJ & 28 GB & 19h45 & 18481 kJ & 28GB  & 0.61  & 1520 \\
    \end{tabular}}
  \label{tab1}
\end{table}

\section{Conclusion and perspectives}

Despite a challenging medical problematic of PD progression at early stages, we have observed that energy and memory efficient methods could yield interesting and comparable results with other studies performed on the same database \cite{pinon2023,Pinon2021} and with similar MR modalities
\cite{Schwarz2013,Du2011,Munoz2019,Munoz2021}. An interesting future work would  be to investigate the possibility to use more structured observations, such as patch-based features \cite{Pinon2021} or latent representations from a preliminary pretext task, provided the task cost is reasonable.
Overall, we have illustrated that the constraints of Green AI  \cite{Schwartz2020} could be considered in medical imaging by producing innovative results  without increasing computational cost or even reducing it.
We have investigated statistical mixture models for an UAD task and shown that their expressivity could account for multivariate reference models, and their much simpler structure made them more amenable to efficient learning than most ANN solutions. Although very preliminary, we hope this attempt will open the way to the development of more methods that can balance the environmental impact of growing energy cost  with the obtained healthcare benefits.

\section{Data use declaration and acknowledgement}

\noindent G. Oudoumanessah was financially supported by the AURA region. This work has been partially supported by MIAI@Grenoble Alpes (ANR-19-P3IA-0003), and was granted access to the HPC resources of IDRIS under the allocation 2022-AD011013867 made by GENCI. The data used in the preparation of this article were obtained from the Parkinson’s Progression Markers Initiative database \href{www.ppmi-info.org/access-data-specimens/download-data}{www.ppmi-info.org/access-data-specimens/download-data} openly available for researchers. 

\newpage

\bibliographystyle{splncs04}
\bibliography{biblio3191}


\section*{Supplementary material}

# Supplementary Material: Towards frugal unsupervised detection of subtle abnormalities in medical imaging

## 1 Weight-based proximity measure for Gaussian mixtures

For a Gaussian distribution with mean $\boldsymbol{\mu}$ and covariance $\boldsymbol{\Sigma} = \mathbf{D}^T \mathbf{A} \mathbf{D}$, the Mahalanobis distance of a sample $\mathbf{y}$ is $(\mathbf{y} - \boldsymbol{\mu})^\top \boldsymbol{\Sigma}^{-1} (\mathbf{y} - \boldsymbol{\mu})$ which can be decomposed as $\sum_{m=1:M} (\mathbf{d}_m^\top (\mathbf{y} - \boldsymbol{\mu}))^2 / A_m$. Each term in the sum can be interpreted as a dimension-wise distance and its inverse as a weight. This inverse corresponds to expression $u_{im}$ in Section 3 of the manuscript with $\nu_m = 0$. Using this similarity we define a proximity for Gaussian mixtures as the maximum over m of these weights whose expressions are the same than for MST mixtures with $\nu_m = 0$.

## 2 Online MST mixture EM updates

We first specify the expression of $\bar{\boldsymbol{\theta}}(\mathbf{s})$ for a single MST distribution. We can define $\bar{\boldsymbol{\theta}}(\mathbf{s})$ as the root of $\mathbf{J}_{\boldsymbol{\phi}}(\boldsymbol{\theta})\mathbf{s} - \frac{\partial \psi}{\partial \boldsymbol{\theta}}(\boldsymbol{\theta}) = \mathbf{0}$, where $\mathbf{J}_{\boldsymbol{\phi}}(\boldsymbol{\theta}) = \partial \boldsymbol{\phi} / \partial \boldsymbol{\theta}$ is the Jacobian of $\boldsymbol{\phi}$, with respect to $\boldsymbol{\theta}$, as a function of $\boldsymbol{\theta}$. Parameters $(\boldsymbol{\mu}, \mathbf{A}, \mathbf{D})$ and $\boldsymbol{\nu}$ can be optimized separately. It follows that,

$\boldsymbol{\mu} = \left(\sum_{m=1}^{M} \frac{s_{3m}}{A_m} (\mathbf{d}_m \mathbf{d}_m^T)\right)^{-1} \left(\sum_{m=1}^{M} \frac{\mathbf{d}_m \mathbf{d}_m^T}{A_m} \mathbf{s}_{1m}\right) = \mathbf{D} \mathbf{S}_3^{-1} \mathbf{v}$ where $\mathbf{S}_3 = diag(s_{31}, \dots s_{3M})$ and $\mathbf{v}$ is defined as $\mathbf{v}^T = (\mathbf{d}_1^T \mathbf{s}_{11}, \dots, \mathbf{d}_M^T \mathbf{s}_{1M})$. For matrix $\mathbf{A}$, we get for each $m$,

$$
\begin{aligned}
A_m &= vec(\mathbf{S}_{2m})^T vec(\mathbf{d}_m \mathbf{d}_m^T) + s_{3m} vec(\mathbf{d}_m \mathbf{d}_m^T)^T vec(\boldsymbol{\mu}\boldsymbol{\mu}^T) - 2 \mathbf{s}_{1m}^T \mathbf{d}_m \mathbf{d}_m^T \boldsymbol{\mu} \\
&= \mathbf{d}_m^T \mathbf{S}_{2m} \mathbf{d}_m - \frac{(\mathbf{d}_m^T \mathbf{s}_{1m})^2}{s_{3m}},
\end{aligned} \tag{1}
$$

where the last equality is obtained by replacing $\boldsymbol{\mu}$ by its expression above. Then, pluging-in the expressions of $\boldsymbol{\mu}$ and $\mathbf{A}$ above and omitting parts that depend on $\boldsymbol{\nu}$ only, it comes,

$$
\mathbf{D} = \arg\min_{\mathbf{D}, \mathbf{D}^T \mathbf{D} = Id} \sum_{m=1}^{M} \log\left(\mathbf{d}_m^T (\mathbf{S}_{2m} - \frac{\mathbf{s}_{1m} \mathbf{s}_{1m}^T}{s_{3m}}) \mathbf{d}_m\right) \tag{2}
$$

With the orthogonality constraint on $\mathbf{D}$, this can be solved using a version of the conjugate gradient algorithm designed for manifolds with the Polak-Ribiere line search computing the Stiefel gradient. See [1, 5] for details. For $\boldsymbol{\nu}$, as for the standard $t$-distribution, we have to solve the following equation in $\nu_m$ for each $m$, $s_{4m} - s_{3m} - \Psi^{(0)}(\nu_m/2) + \log(\nu_m/2) + 1 = 0$, where $\Psi^{(0)}$ is the digamma function.

Then, the online EM for a $K$-component mixture $\mathcal{M}$ can be derived from the online EM for a single component. It consists of computing $\bar{\boldsymbol{\theta}}_{\mathcal{M}}(\mathbf{s}_{\mathcal{M}})$ where $\mathbf{s}_{\mathcal{M}} = (s_{01}, \mathbf{s}_{\mathcal{M}1}^\top, \dots, s_{0K}, \mathbf{s}_{\mathcal{M}K}^\top)$ and $\mathbf{s}_{\mathcal{M}k}$ has the structure given in the manuscript for one MST distribution. We can show [2], that $\bar{\boldsymbol{\theta}}_{\mathcal{M}}(\mathbf{s}_{\mathcal{M}})^\top = (\bar{\pi}_1(\mathbf{s}_{\mathcal{M}}), \bar{\boldsymbol{\theta}}(\mathbf{s}_{\mathcal{M}1}/s_{01})^\top, \dots, \bar{\pi}_K(\mathbf{s}_{\mathcal{M}}), \bar{\boldsymbol{\theta}}(\mathbf{s}_{\mathcal{M}K}/s_{0K})^\top)$ where $\bar{\pi}_k(\mathbf{s}_{\mathcal{M}}) = s_{0k}/(\sum_{z=1:K} s_{0z})$ and $\bar{\boldsymbol{\theta}}$ is the expression found for one single MST component, detailed in the manuscript.

# 3 Improved implementation of online Gaussian mixture EM

We cannot detail the online EM for K-Gaussian mixtures but the algorithm can be found in previous work. We focus here on our proposed improvements. We use the notation in [3] just changing the indices to match notation in our previous section, that is for each $k = 1 : K$, we set $(s_{0k}, \mathbf{s}_{1k}, \mathbf{S}_{2k}) = (\mathbf{s}_{1k}, \mathbf{s}_{2k}, \mathbf{S}_{3k})$. At iteration $i$, the update of $\boldsymbol{\theta}^{(i)}$ involves the update of $K$ covariance matrices, for $k = 1 : K$, $\boldsymbol{\Sigma}_k^{(i)} = \mathbf{S}_{2k}^{(i)}/s_{0k}^{(i)} - \mathbf{s}_{1k}^{(i)}\mathbf{s}_{1k}^{(i)T}/s_{0k}^{(i)2}$. The subsequent update of $\mathbf{s}^{(i+1)}$ then requires the inverse covariances (i.e. precisions) and their determinants whose computation is costly. We thus propose to work directly with precision matrices. Applying the Sherman-Morrison formula [4] to the equation above gives,

$$\boldsymbol{\Sigma}_k^{(i)-1} = s_{0k}^{(i)}\mathbf{S}_{2k}^{(i)-1} + \frac{\mathbf{S}_{2k}^{(i)-1}\mathbf{s}_{1k}^{(i)}\mathbf{s}_{1k}^{(i)T}\mathbf{S}_{2k}^{(i)-1}}{1 - \frac{1}{s_{0k}^{(i)}}\mathbf{s}_{1k}^{(i)T}\mathbf{S}_{2k}^{(i)-1}\mathbf{s}_{1k}^{(i)}}.$$

To avoid any matrix inversion, we also apply the Sherman-Morrison formula to the update of $\mathbf{S}_2$,

$$\mathbf{S}_{2k}^{(i)-1} = \frac{1}{1-\gamma_i}\mathbf{S}_{2k}^{(i-1)-1} - \frac{\gamma_i\tau_{ki}}{(1-\gamma_i)^2}\frac{\mathbf{S}_{2k}^{(i-1)-1}\mathbf{y}_i\mathbf{y}_i^T\mathbf{S}_{2k}^{(i-1)-1}}{1 + \frac{\gamma_i\tau_{ki}}{(1-\gamma_i)}\mathbf{y}_i^T\mathbf{S}_{2k}^{(i)-1}\mathbf{y}_i}.$$

Similarly, the determinant version of the Sherman-Morrison formula gives our new updates for the precision determinants and the $\mathbf{S}_2$ statisics:

$$\det(\boldsymbol{\Sigma}_k^{(i)-1}) = \frac{s_{0k}^{(i)M}}{1 - \frac{\mathbf{s}_{1k}^{(i)T}\mathbf{S}_{2k}^{(i)-1}\mathbf{s}_{1k}^{(i)}}{s_{0k}^{(i)}}}\det(\mathbf{S}_{2k}^{(i)-1})$$

$$\det(\mathbf{S}_{2k}^{(i)-1}) = \left(1 + \frac{\gamma_i\tau_{ik}}{1-\gamma_i}\mathbf{y}_i^T\mathbf{S}_{2k}^{(i-1)-1}\mathbf{y}_i\right)^{-1}(1-\gamma_i)^{-M}\det(\mathbf{S}_{2k}^{(i-1)-1}).$$

A similar improvement can be derived for a mini-batch online EM [3] by applying both Sherman-Morrison formulas recursively.


\end{document}